\documentclass[fleqn,usenatbib]{mnras}
\usepackage{newtxtext,newtxmath}
\usepackage[T1]{fontenc}
\DeclareRobustCommand{\VAN}[3]{#2}
\let\VANthebibliography\thebibliography
\def\thebibliography{\DeclareRobustCommand{\VAN}[3]{##3}\VANthebibliography}

\usepackage{graphicx}	% Including figure files
\usepackage{amsmath}	% Advanced maths commands

\newcommand{\obj}{SDSS\,J0826+5630} 
\newcommand{\msun}{M$_\odot$}

\newcommand{\hi}{{\sc H\,i}}

\newcommand{\Mhi}{$M_{\rm HI}$}

\title[Veracity of a claimed $z=1.3$ \hi\ detection]{On the implausible physical implications of a claimed lensed neutral hydrogen detection at redshift $z = 1.3$ }
\author[Deane et al.]{Roger P. Deane$^{1,2}$, Tariq Blecher$^{3,4}$, Danail Obreschkow$^{5, 6}$, Ian Heywood$^{7,3,4}$\\
$^1$ Wits Centre for Astrophysics, University of the Witwatersrand, 1 Jan Smuts Avenue, 2000, Johannesburg, South Africa\\
$^2$ Department of Physics, University of Pretoria, Hatfield, Pretoria, 0028, South Africa \\
$^3$ Centre for Radio Astronomy Techniques and Technologies, Department of Physics and Electronics, Rhodes University, Makhanda 6140, South Africa\\
$^4$ South African Radio Astronomical Observatory, 2 Fir Street, Observatory, 7925, South Africa \\
$^5$ International Centre for Radio Astronomy Research (ICRAR), M468, University of Western Australia, WA 6009, Australia\\
$^6$ ARC Centre of Excellence for All Sky Astrophysics in 3 Dimensions (ASTRO 3D)\\
$^7$ Astrophysics, University of Oxford, Denys Wilkinson Building, Keble Road, Oxford, OX1 3RH, UK 
}
\date{Accepted 2024 September 09. Received 2024 September 09; in original form 2024 May 07}
\begin{document}
\label{firstpage}
\pagerange{\pageref{firstpage}--\pageref{lastpage}}
\maketitle

\begin{abstract}
The Square Kilometre Array mid-frequency array will enable high-redshift detections of neutral hydrogen (\hi) emission in galaxies, providing important constraints on the evolution of cold gas in galaxies over cosmic time. Strong gravitational lensing will push back the \hi\ emission frontier towards cosmic noon ($z\sim2$), as has been done for all prominent spectral lines in the interstellar medium of galaxies. Chakraborty \& Roy (2023, MNRAS, 519, 4074)  report a $z=1.3$ \hi\ emission detection towards the well-modelled, galaxy-scale gravitational lens, SDSS\,J0826+5630. We carry out \hi\ source modelling of the system and find that their claimed \hi\ magnification, $\mu_{\rm HI} = 29 \pm 6$, requires an \hi\ disk radius of $\lesssim 1.5$~kpc, which implies an implausible mean \hi\ surface mass density in excess of $\Sigma_{\rm HI} > 2000 $~\msun\,pc$^{-2}$. This is several orders of magnitude above the highest measured peak values ($\Sigma_{\rm HI} \sim  10~{\rm M}_\odot\,{\rm pc}^{-2}$), above which \hi\ is converted into molecular hydrogen. Our re-analysis requires this to be the highest \hi\ mass galaxy known (\Mhi$~\sim 10^{11}$~\msun), as well as strongly lensed, the latter having a typical probability of order 1 in 10$^{3-4}$. We conclude that the claimed detection is spurious. 
\end{abstract}


%%%%%%%%%%%%%%%%%%%%%%%%%%%%%%%%%%%%%%%%%%%%%%%%%%

%%%%%%%%%%%%%%%%% BODY OF PAPER %%%%%%%%%%%%%%%%%%

\section{Introduction}
Gravitationally lensed neutral hydrogen and hydroxyl at intermediate to high redshift is an exciting area to be explored by the Square Kilometre Array (SKA) and its precursors/pathfinders \citep[][]{Deane2015,Deane2016,Hunt2016,Blecher2019,Ranchod2022,Button2024}. This will enable the study of individual objects at higher redshift to compare and contrast with statistical methods \citep[e.g. stacking, intensity mapping; ][]{Bera2019,Chowdury2020,Cunnington2023} and \hi\ absorption \citep[e.g.][]{Deka2023}, as well as hydrodynamical simulations \citep[e.g.][]{Obreschkow2009,Lagos2011,Dave2020}. Direct, unbiased measures of the \hi\ emission in individual galaxies are necessary to complement and cross-check statistical measures with large samples, given the important role of \hi\ as a transitional cold gas phase in the baryon cycle \citep{Dave2012,Lilly2013}. Furthermore, spatially-resolved \hi\ disk dynamics will provide key constraints to fundamental galaxy evolution properties, including disk angular momentum \citep{Obreschkow2014_momentum}, particularly since the \hi\ reservoir extends to larger radii than other tracers of the disk kinematics. However, doing so at cosmologically significant distances remains a challenge, due to the intrinsic faintness of the line. Gravitational lensing has been suggested as a tool to expand to \hi\ emission frontier to redshifts comparable to other important spectral lines that stem from the interstellar medium of galaxies \citep{Deane2015}. Magnification of \hi\ disks will enable spatially-resolved observations not otherwise possible with a given instrument.

\citet[][CR23 hereafter]{CR2023} present upgraded Giant Metrewave Radio Telescope (uGMRT) Band 4 observations ranging between  550-650~MHz of the strongly lensed source \obj. The observations were carried out under Proposal 34$\_$066 (Principal Investigator: T. Blecher). The $z=1.3$ source is located at RA = 08$^{\rm h}$26$^{\rm m}$39.858$^{\rm s}$, Dec = +56$^\circ$30'35.97'' and all further relevant observational details are described in CR23. The current paper concerns itself with the lens modelling and derived physical properties reported in CR23.

The discovery of \obj\ was presented in \citet{Shu2017} in the Sloan Lens ACS (SLACS, \citealt{Bolton2008}) Survey for the Masses (S4TM) Survey. They performed follow-up {\sl Hubble Space Telescope (HST)} imaging of 118 lens candidates as part of a low-mass extension to SLACS, confirming 40 of these as bona fide strong lenses. \obj\ is one of the lower mass lenses in the confirmed sample (stellar velocity dispersion, $\sigma_v =163 \pm 8$~km\,s$^{-1}$); however, it also has one of the highest reported optical magnifications ($\mu_{\rm opt} \sim 105$). The observations approved under GMRT Cycle-34 programme 34$\_$066 were motivated to search for strongly lensed \hi\ galaxies accessible with the (then) newly available uGMRT backend. 

\begin{figure*}
	\includegraphics[width=0.99\textwidth]{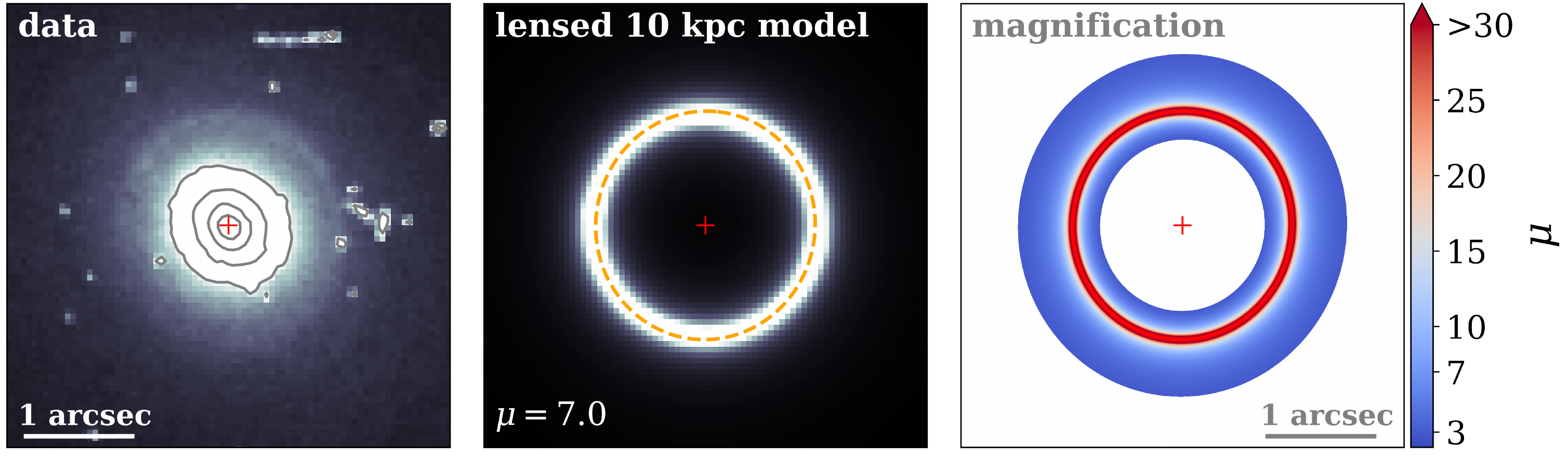}
    \caption{Left: { \sl HST} F814W image of the \obj\ system showing foreground lens at the centre (grey contours and red cross) and fainter $\sim$1-arcsec radius Einstein ring, with the colour scale saturated for enhanced contrast. The red cross corresponds to the lens centre. Middle: example $r_{\rm s} = 10$~kpc lensed exponential source centred on the lens to form a full Einstein ring and demonstrate the lensing configuration. The dashed orange ellipsoid is the critical curve. The total magnification for this 10~kpc source is $\mu = 7.0$, a factor of $\sim4$ lower than the \hi\ magnification reported in CR23. Right: Image-plane magnification, which illustrates how small the image-plane region is that is magnified by a factor larger than $\mu \gtrsim 30$. All three images are $4 \times 4$~arcsec in extent.}
    \label{fig:datamoderesid}
\end{figure*}

More recently, CR23 analyze these data and present three main results: (1) a claimed 5$\sigma$ detection of gravitationally lensed neutral hydrogen in this $z=1.3$ system; (2) a measured \hi-to-stellar mass ratio of $2.363 \pm 0.14$; and (3) the measurement of a spatially resolved \hi\ component with a $\sim$6.8~arcsec major axis which is tangential to the critical curve of a 1~arcsec Einstein radius lens model. These results would be a remarkable achievement as the source is at a redshift several factors larger than the current highest redshift \hi\ emission detections. 

The claimed \hi\ mass of $M_{\rm HI} = 0.9 \times 10^{10} \ {\rm M}_\odot$ implies an \hi\ diameter of $\sim$55~kpc, according to the \citet{Wang2016} relation imposed by CR23. This corresponds to an {\sl unlensed} angular extent of 6.5~arcsec at $z=1.3$. As we will demonstrate in this Letter, this is highly inconsistent with the size of the region ($\lesssim0.1$~arcsec) in which their claimed \hi\ magnification factor of $\sim30$ is possible. Therefore, something must be incorrect in the CR23 analysis. In our reanalysis, we show that even if adherence to the \citet{Wang2016} relation is relaxed, the implications of the claimed CR23 detection remain implausible. In Sec.~\ref{sec:methods}, we describe the lens model and Monte Carlo ray-tracing simulations, followed by the implied physical parameters and a conclusion in Sec.~\ref{sec:conclusion}. We assume Planck 2018 cosmological parameters throughout the paper \citep{Planck2018}. 

\section{Method and Discussion}\label{sec:methods}

\subsection{Lens Model}

In \citet{Blecher2019}, we develop a method to model lensed \hi\ sources, incorporating realistic radial profiles \citep{Obreschkow2009} and enforcing the \hi\ mass-diameter relation \citep{Wang2016}. CR23 state they use this exact same method, but with peculiar results that motivate us to re-analyse the \obj\ system. For our modelling, we use the same {\sl HST}-derived lens model presented in \citet{Shu2017}, just as CR23 do. We can, therefore, be confident that conceptually, the modelling approaches of both lens and source are identical, and any significant differences must be an incorrect implementation thereof. 

The \citet{Shu2017} derived lens model finds a Single Isothermal Ellipsoid (SIE) mass-density profile (with no external shear) as sufficient to accurately model the system. The derived SIE model for \obj\ has an Einstein radius $\theta_{\rm E} = 1.01$~arcsec, which is fractionally larger than that implied by the SDSS-measured stellar velocity dispersion of $\sigma = 163\pm8$~km\,s$^{-1}$ for a foreground lens at $z_{\rm l} = 0.1318$ and a background source at $z_{\rm s} = 1.2907$. The SIE minor-to-major axis ratio is $q = 0.96$ at a position angle of 82~deg. The lens is designated as Grade `A', meaning it has convincing signs of multiple imaging, which is clear from the near-circular Einstein ring seen in Fig.~\ref{fig:datamoderesid} (left panel). The foreground source morphology shows a smooth, well-modelled elliptical galaxy light distribution. All of these traits make this a comparatively straightforward, unambiguous system to model, with the optical source model requiring just a single S\'ersic component to achieve a reduced $\chi_\nu^2$ goodness-of-fit value of 1.7.

\subsection{Monte Carlo Simulations}

The primary aims of our probabilistic analysis are the following. 

\begin{enumerate}
    \item Derive probability distributions of the \hi\ magnification factor using \citet{Blecher2019} for direct comparison with the \hi\ magnification posterior probability distribution presented in CR23.
    \item Constrain the source-plane \hi\ emission size (which is not reported in CR23) in order to compute the implied mean \hi\ mass surface density, $\Sigma_{\rm HI}$. 
    \item Compute the implied intrinsic \hi\ mass and compare it to the highest known \hi\ masses. 
\end{enumerate}

To this end, we perform a Monte Carlo simulation using the {\sc lenstronomy}\footnote{https://github.com/lenstronomy/lenstronomy} package \citep{Birrer2018,Birrer2021}, which has a large development group, user base and has been extensively tested and validated for a number of lensing applications, ranging from non-parametric source reconstruction, dark matter substructure constraints, large-scale survey simulations, and Hubble parameter estimation using time delays. We use {\sc lenstronomy} to carry out $10^{5.3}$ ray-tracing realisations (over an order of magnitude more than CR23), varying the parameters presented in \citet{Blecher2019}. These include the \hi\ mass (and hence disk size through the \citealt{Wang2016} relation), disk inclination, position angle, and centroid coordinates (and hence the impact parameter between lens and source), and $R^{\rm c}_{\rm mol}$ a parameter related to the ratio of molecular mass, $M_{\rm H_2}$, to atomic hydrogen mass, $M_{\rm HI}$, \citep[see][]{Obreschkow2009},

\begin{equation}
 M_{\rm H_2}/M_{\rm HI} = (3.44 R^{\rm c\ -0.506}_{\rm mol}+4.82R^{\rm c\ -1.054}_{\rm mol})^{-1}.
\end{equation}

As in \citet{Obreschkow2009}, \citet{Blecher2019}, and CR23,  we adopt the axisymmetric \hi\ mass surface density model of \citet{Obreschkow2009},

\begin{equation}
\Sigma_{\rm HI} (r) = \frac{M_{\rm  H}/(2\pi r_{\rm disk}) \exp{(-r/r_{\rm disk})}}{1+R^{\rm c}_{\rm mol}\exp{(-1.6 r/r_{\rm disk})}},
\label{eq:HI_profile}
\end{equation}

\noindent where $r$ is the galactocentric radius in the disk plane, $ M_{\rm  H} = M_{\rm H_2}+M_{\rm HI}$,  $r_{\rm disk}$ is the scale length of the neutral hydrogen disk (atomic plus molecular). This \hi\ radius, $r_{\rm disk}$, is used throughout this Letter, unless explicitly stated otherwise. As a reference, $r_{\rm disk}$ is approximately one-third of the commonly used radius at which the \hi\ mass surface density falls below 1~M$_\odot$\,pc$^{-2}$, for the distribution of $R^{\rm c}_{\rm mol}$ used in this work. 

Following tests with a much larger parameter space, we limit the parameter space explored to a range consistent with that stated in CR23 (which are all repeated exactly from \citealt{Blecher2019}). We employ a uniformly-drawn distribution for the \hi\ centroid impact parameter = [0, 0.5]~arcsec; $\log(M_{\rm HI}/{\rm M}_\odot)$ = [7, 12] (which sets the \hi\ diameter via the \citealt{Wang2016} relation); disk inclination range = [0, $\frac{\pi}{2}$], disk position angle = [0, $\pi$]. The $R^{\rm c}_{\rm mol}$ parameter is drawn from a log-normal distribution with mean and standard deviation of -1 and 0.3, respectively. This log-prior was employed in the analysis of a $z\sim0.4$ galaxy in \citet{Blecher2019}, given its stellar mass and local universe results from a sample with measured atomic and molecular hydrogen \citep{Catinella2018}. CR23 do not justify its use for a $z=1.3$ galaxy, despite a wealth of observational and simulated results that suggest significant evolution at $z > 1$. Nonetheless, to assess their model, we repeat our analysis with this same log-prior, even if it is not justified for \obj. The lens and source modelling Monte Carlo simulations are therefore precisely configured to exactly what CR23 describe, following \citet{Blecher2019}, with the only difference being that we do not include \hi\ masses between $6 < \log(M_{\rm HI}) < 7$ to the simulation, which adds unnecessary processing time with negligible changes to the results. 

\begin{figure}
	\includegraphics[width=0.5\textwidth]{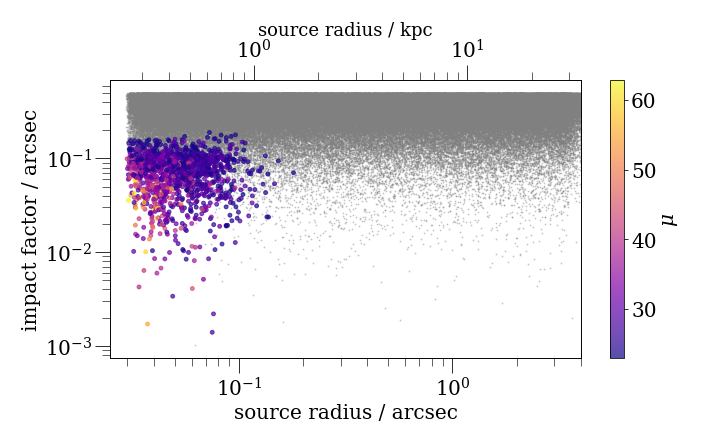}
    \caption{Results of the Monte Carlo analysis with 10$^{5.3}$ realisations (all shown in grey). Realisations with an \hi\ magnification larger than $\mu > 23$ (i.e. the mean minus one standard deviation quoted in CR23) are colourised. These demonstrate that \hi\ disk radii $\lesssim 1.5$~kpc and impact parameters of $\lesssim 0.15$~arcsec are required to achieve magnifications statistically consistent with that quoted in CR23 using the \citet{Shu2017} lens model.}
    \label{fig:MonteCarloMu}
\end{figure}

From this Monte Carlo analysis, we isolate those positions and source radii that have $\mu_{\rm HI} \geq 23$ (i.e. the CR23 quoted mean \hi\ magnification \emph{minus} one standard deviation), which conservatively ought to include the same position that CR23 claim to have fixed their \hi\ disk to. We can now assume we have isolated the small region of the parameter space using the \citet{Shu2017} lens model where it is possible to achieve magnifications consistent with their claimed $\mu_{\rm HI}$. In Fig.~\ref{fig:MonteCarloMu}, we show the results of this analysis, identifying the parameter space that corresponds to magnifications greater than $\mu > 23$. This demonstrates that magnifications consistent with that quoted in CR23 require \hi\ disk radii $\lesssim 1.5$~kpc. 

\subsection{Implied \hi\ surface mass density}

\begin{figure}
	\includegraphics[width=0.47\textwidth]{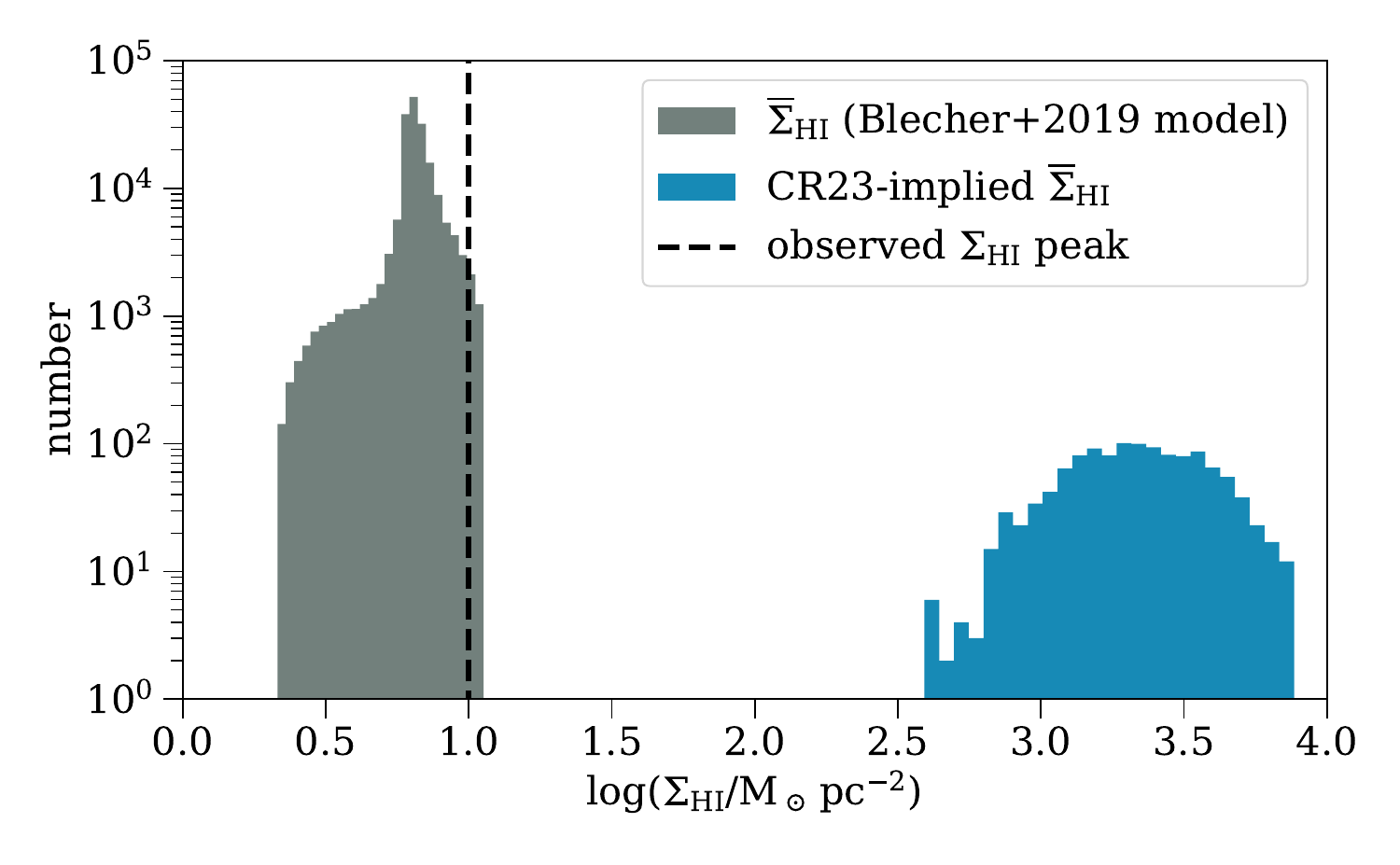}
    \caption{Implied mean \hi\ mass surface density, $\Sigma_{\rm HI}$ (grey), within $r_{\rm disk}$ for all Monte Carlo realisations using the \citet{Blecher2019} source model. The black dashed vertical line indicates the typical peak \hi\ mass surface density measured in the local universe with sub-kpc spatial resolution \citep[e.g.][]{Bigiel2008}. In blue, we show the implied $\Sigma_{\rm HI}$ distribution, if one assumes the quoted CR23 \hi\ flux density and only considers those realisations with magnifications $\mu > 23$ (i.e. within 1 standard deviation of their quoted magnification). This demonstrates that the CR23 implementation of the \citet{Blecher2019} model and claimed \hi\ detection cannot both be correct. }
    \label{fig:SigmaHI}
\end{figure}

CR23 present what they claim is a posterior probability distribution of $\mu_{\rm HI}$, with a quoted mean and standard deviation of $\mu_{\rm HI} = 29.37\pm 6$. In Fig.~\ref{fig:SigmaHI} we show the corresponding distributions of {\sl mean} intrinsic \hi\ mass surface densities within $r_{\rm disk}$ for all realisations that have magnifications of $\mu > 23$. What is overwhelmingly clear is this distribution is 2-3 orders of magnitude higher than any measured {\sl peak} \hi\ mass surface density to date \citep[e.g.][]{Bigiel2008}. The \citet{Blecher2019} model enforces the \citet{Wang2016} relation, and so naturally one can find its implied \hi\ mass surface densities within $r_{\rm disk}$ near this peak, as seen in Fig.~\ref{fig:SigmaHI}. 

 As is supported in the discussion below, it is highly implausible that a considerable fraction of a galaxy's \hi\ reservoir would reach mass surface densities much above $\Sigma_{\rm HI} \gtrsim 10~{\rm M}_\odot\,{\rm pc}^{-2}$ at any time. This appears to be a universal saturation point supported by a wide range of observations. Furthermore, cosmological hydrodynamical zoom-in simulations of evolving galaxies with sub-grid models for hydrogen gas phases, accounting for density, temperature, metallicity and UV background \citep{Wang2018}, show the evolution of galaxy sizes and atomic gas fractions that are in line with a saturation of order $\Sigma_{\rm HI} \sim 10~{\rm M}_\odot\,{\rm pc}^{-2}$ at all cosmic times. The strongest observational test we can likely consider at this point is the $\Sigma_{\rm HI}$ in the most intense starbursts in the local universe, often undergoing mergers, which serve as high-redshift analogues in many senses. Even these sites of extreme molecular gas densities do not show significantly super-critical \hi\ surface densities \citep[e.g.][]{vanderHulst1979,Jackson1987}. Furthermore, physical models of H$_2$ formation suggest that even a strong dissociating UV background (an order of magnitude larger than the Milky Way level) paired with low dust fractions (10 per cent of the Milky Way level) would not suppress the H$_2$ formation enough to increase the \hi\ saturation density by two orders of magnitude \citep{Gnedin2011}.

This result alone renders the CR23 claims to be implausible. 

\subsection{Astrometry and tangential magnification}

CR23 also report that a component in the central channel map is spatially resolved in a tangential direction to the assumed critical curve, with a fitted image-plane major axis of $\theta_{\rm maj} = 6.8$~arcsec (no uncertainty stated) at a position angle of $52.2 \pm 9$~deg, convolved with a PSF FWHM of $13.37 \times 12.32$ arcsec, PA = 37~deg. This assumed (but not demonstrated) tangential magnification is used in support of the detection's veracity. 

The astrometric uncertainty of the claimed central channel \hi\ component is $\sigma_{\rm pos} = \theta_{\rm maj}/{\rm SNR} \sim 13/4 \gtrsim 3$~arcsec. This uncertainty is a factor of $\gtrsim3$ larger than the Einstein radius, making the \hi\ source centroid statistically consistent with any position angle from the lens centroid, and therefore, this claimed detection could be subjected to tangential broadening in any direction. This nullifies its use as a supporting argument for the claimed detection in CR23. We note that since the source is spatially resolved, the positional uncertainty is even larger than that what we estimate here. Furthermore, the absolute astrometry of both the uGMRT and {\sl HST} are neither compared nor incorporated. Since CR23 do not quote the relevant positional uncertainties, we consider our estimate as a conservative lower limit.

\subsection{Spatial scales summary}

\begin{figure}
	\includegraphics[width=0.47\textwidth]{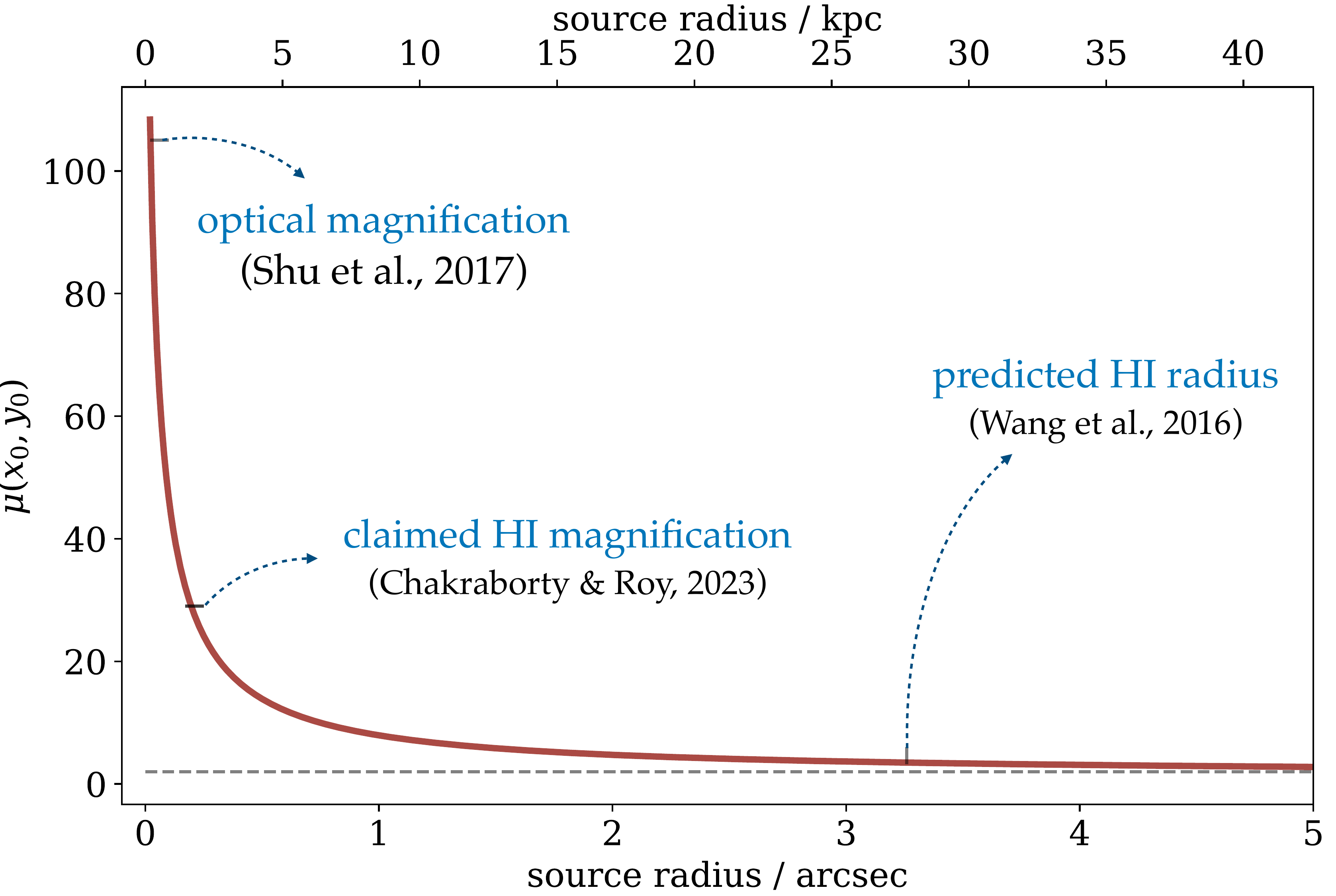}
    \caption{Magnification profile for a face-on exponential disk centred on the lens. This recovers the high optical magnification ($\mu_{\rm opt} \sim 105$) reported in \citet{Shu2017}. The radius ($r = 0.17$~arcsec) required to achieve a magnification equal to the CR23 value ($\mu = 29$) using the \citet{Shu2017} model in indicated. Also shown is the radius beyond which the implied mean \hi\ mass surface density falls below $\Sigma_{\rm HI} = 10~{\rm M}_\odot\,{\rm pc}^{-2}$, the peak value measured in the local universe with sub-kpc resolution \citep{Bigiel2008}. The black dashed horizontal line corresponds to the formal strong lensing criterion (i.e. $\mu = 2$). }
    \label{fig:magn_vs_radius}
\end{figure}

In Fig.~\ref{fig:magn_vs_radius}, we summarize the results for relevant spatial scales and their corresponding magnifications. This assumes a face-on exponential disk centred on the lens, which is consistent with the Monte Carlo analysis presented in Fig.~\ref{fig:MonteCarloMu} and reproduces the \citet{Shu2017} derived optical magnification of $\mu = 105$, seen at the top-left of the figure. This is further justified by the narrow linewidth (FWHM $\lesssim 72$~km\,s$^{-1}$) for an expected \hi\ massive galaxy. CR23 also note they find no dependence on disk position angle, inclination and the $R^{\rm c}_{\rm mol}$ parameter. Also annotated in Fig.~\ref{fig:magn_vs_radius} is the CR23 magnification at an indicative disk radius of $\lesssim 0.1$~arcsec, as well as the \citet{Wang2016} predicted radius ($r = 3.26$~arcsec), beyond which the implied mean \hi\ mass surface density falls below the peak values measured in the local universe ($\Sigma_{\rm HI} = 10~{\rm M}_\odot\,{\rm pc}^{-2}$). 

\subsection{Implied intrinsic \hi\ mass}

Finally, we turn to the implied \hi\ mass for a more realistic $\mu_{\rm HI}$ distribution. Clearly the mean $\mu_{\rm HI}$ must be substantially lower than the $\mu_{\rm HI} = 29.37 \pm6$ claimed in CR23, which in turn implies the intrinsic \hi\ mass must be significantly higher than $M_{\rm HI} = 0.9 \times 10^{10}$~M$_\odot$. As has been demonstrated with the implausible \hi\ mass surface densities, the CR23 results lie several orders of magnitude off of this relation. 

\begin{figure}
	\includegraphics[width=0.47\textwidth]{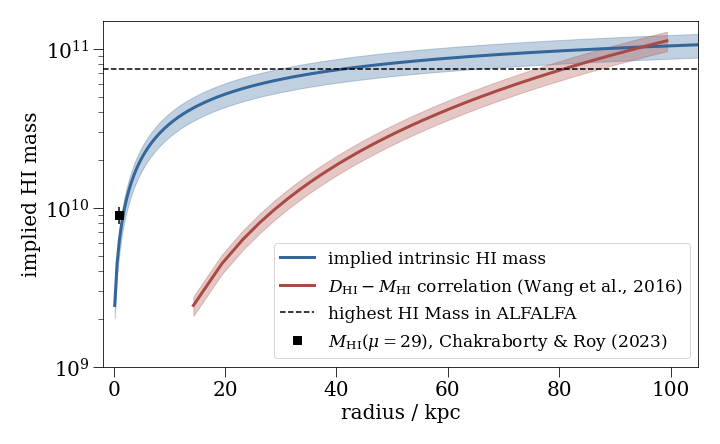}
    \caption{Implied intrinsic \hi\ mass and quoted uncertainty (blue curve and shading) based on the reported CR23 apparent \hi\ mass (i.e. claimed intrinsic mass multiplied by their magnification of 29) and the implementation of the \citet{Shu2017} lens model in this work. The red curve and shading correspond to the  $M_{\rm HI}-D_{\rm HI}$ \citet{Wang2016} relation and scatter (which the \citealt{Blecher2019} method enforces). These two curves only intersect at a radius $r > 90$~kpc, corresponding to magnifications of $\mu < 3$ and $M_{\rm HI}$ larger than the highest \hi\ masses recorded in the ALFALFA survey \citep[dashed horizontal line][]{Jones2018}. The CR23 \hi\ mass and magnification are indicated by the black square.  }
    \label{fig:impliedHImass}
\end{figure}

Using the magnification profile from Fig.~\ref{fig:magn_vs_radius}, we calculate the implied \hi\ mass based on the CR23 \hi\ flux density and more realistic \hi\ magnifications. This is shown as the blue line in Fig.~\ref{fig:impliedHImass}, which is consistent with the CR23 value (black square) for an \hi\ disk radius of $r_{\rm disk} \sim 1.5$~kpc. For comparison, we show the \hi\ mass-diameter, $M_{\rm HI}-D_{\rm HI}$, relation measured in the low-redshift universe \citep{Wang2016}. The two curves intersect at an \hi\ mass of $\gtrsim 10^{11}$~M$_\odot$ and an \hi\ disk radius of $\sim 100$~kpc, corresponding to a magnification of $\mu \lesssim 2.5$. Also shown is the highest \hi\ mass in the ALFALFA survey \citep[$M_{\rm HI} \sim 7.4 \times 10^{10}$~M$_\odot$, ][]{Jones2018}, which lies below this intersection point. Significant evolution of the \hi\ mass function and $\Omega_{\rm HI}$ are not supported by current observational constraints \citep[e.g.][]{Walter2020} and a comparison of a suite of state-of-the-art cosmological hydrodynamical simluations \citep{Dave2020}. This is due, in part, to the very $\Sigma_{\rm HI}$ saturation level that the CR23 violates by 2-3 dex. Therefore, significant evolution of \hi\ in galaxies does not appear to be a possible solution to the uncomfortably high \hi\ mass implied, particularly if this outlier is to be gravitational lensed as well. While the \citep{Wang2016} relation may not be accurate at $z = 1.3$, there is no evidence at present from both observations and cosmological hydrodynamical simulations for significant deviation ($\gg 1$~dex) from the $z=0$ relation.

Irrespective of the noted rarity of \hi\ masses as large as those implied by the CR23 results, if they had indeed followed \citet{Blecher2019} as they claim they did, the \hi\ source radius would be coupled to the intrinsic \hi\ mass, which would be $\sim28$~kpc, not $\lesssim 1.5$~kpc. This discrepancy in the \hi\ source-plane size casts major doubt as to the correct implementation of the CR23 \hi\ source modelling method. As can be seen in Fig.~\ref{fig:magn_vs_radius}, we expect emission on scales we expect of the \hi\ to have relatively low magnifications ($\mu \ll 29$).

The above arguments render the CR23 claim implausible and point to a false-positive detection. We offer the more prosaic likelihood often seen in spectral line data at GHz frequencies: that this is a relatively frequently observed artefact of imperfect continuum subtraction of a spatially resolved continuum source,  resulting in part from the variable radio frequency interference flagging statistics which are frequency dependent. The resultant changes to the per-channel point spread function can often result in narrow spectral spikes as presented in CR23. Furthermore, ionospheric scintillation could potentially also cause spurious residual spectral (and spatial) features in the vicinity of continuum sources at these frequencies.

\section{Conclusion}\label{sec:conclusion}

We conclude that the CR23 claims are unlikely to be correct for the following reasons. 

\begin{enumerate}
    \item We demonstrate that \hi\ sources with disk radii $\lesssim 1.5$~kpc and low impact factors ($< 0.1$~arcsec) are required to recover magnifications that are consistent within the 68 per cent confidence intervals of the \hi\ magnification posterior probability distribution function presented in CR23. This implies a remarkably small \hi\ disk radius of $\lesssim1$~kpc which is unseen in the low-redshift universe, nor at higher redshifts with cosmological hydrodynamical simulations for \hi\ masses of $M_{\rm HI} \gtrsim 10^{10}$~M$_\odot$.
    \item The large \hi\ masses derived from the CR23 spectrum results in implausible \hi\ mean mass surface densities, 2-3 orders of magnitude larger than the peak values measured in the local universe. This point remains true regardless of whether or not the CR23 magnification posterior is assumed correct. 
    \item  We show that the \hi\ astrometric uncertainty is a factor of $\gtrsim3$ larger than the Einstein radius. This nullifies the CR23 claim that a spatially-resolved 6.8~arcsec component with a major axis perpendicular the critical curve supports the validity of their claimed detection. 
    \item If one adopts the \citet{Shu2017} lens model and a reasonable range of \hi\ sizes (and implied mean surface mass densities), this requires that the source has amongst the highest \hi\ masses ever measured at low redshift and predicted at high redshift by cosmological hydrodynamical simulations. The low probability of strong lensing ($\lesssim 10^{-3}$) and absence of any significant bias toward high \hi\ mass selection in this source's discovery leads to an uncomfortably low probability requirement that it simultaneously be the highest known \hi\ mass in the universe, as well as strongly lensed. 
\end{enumerate}

We conclude that the physical implications of their results, which are not explored in their paper, are implausible, offering a more prosaic possibility that this is a false-positive detection caused by imperfect continuum subtraction of marginally-resolved continuum source with a frequency-variable point spread function.

It is critically important to open the \hi\ emission frontier out to large cosmological distances to compare with statistical  \hi\ detection approaches and place independent, unbiased constraints on the evolution of \hi\ in galaxies over cosmic time. Therefore, claimed detections must be inspected with great scrutiny before utilized in this endeavour. This is an important input into SKA \hi\ survey designs and has broader value for several fields within contemporary astrophysics. Dedicated campaigns with the SKA precursors/pathfinders will continue to push this frontier back until the SKA1-mid begins observations that will undoubtedly revolutionize our view of of the high-redshift \hi\ universe, and our depth of understanding of baryon cycle in galactic halo, with strong lensing predicted to play a major role.

\section*{Acknowledgements}
We thank the anonymous reviewer for helpful comments. RPD, TB, and IH acknowledge support from the South African Radio Astronomy Observatory, which is a facility of the National Research Foundation (NRF), an agency of the Department of Science and Innovation (DSI). RPD's research is funded by the South African Research Chairs Initiative of the DSI/NRF. DO is a recipient of an Australian Research Council Future Fellowship (FT190100083) funded by the Australian Government. Based on observations made with the NASA/ESA Hubble Space Telescope, and obtained from the Hubble Legacy Archive, which is a collaboration between the Space Telescope Science Institute (STScI/NASA), the Space Telescope European Coordinating Facility (ST-ECF/ESA) and the Canadian Astronomy Data Centre (CADC/NRC/CSA). IH acknowledges support of the STFC consolidated grant [ST/S000488/1] and [ST/W000903/1], and from a UKRI Frontiers Research Grant [EP/X026639/1]. IH acknowledges support from Breakthrough Listen. Breakthrough Listen is managed by the Breakthrough Initiatives, sponsored by the Breakthrough Prize Foundation.

%%%%%%%%%%%%%%%%%%%%%%%%%%%%%%%%%%%%%%%%%%%%%%%%%%
\section*{Data Availability}

The data relevant to this Letter and \citet{CR2023} and are publicly available in GMRT archive \url{https://naps.ncra.tifr.res.in/goa}, under proposal code 34$\_$066 (PI: Blecher).



\bibliographystyle{mnras}
\bibliography{refs} % if your bibtex file is called example.bib

%%%%%%%%%%%%%%%%%%%%%%%%%%%%%%%%%%%%%%%%%%%%%%%%%%

%%%%%%%%%%%%%%%%%%%%%%%%%%%%%%%%%%%%%%%%%%%%%%%%%%

% Don't change these lines
\bsp	% typesetting comment
\label{lastpage}
\end{document}